\documentclass[sigconf]{acmart}
\usepackage{algorithm} 
\usepackage{algorithmic}
\usepackage{multirow}
\usepackage{mathrsfs}
\newcommand{\nop}[1]{}
\usepackage{marvosym}
\copyrightyear{2020} 
\acmYear{2020} 
\setcopyright{acmcopyright}\acmConference[SIGIR '20]{Proceedings of the 43rd International ACM SIGIR Conference on Research and Development in Information Retrieval}{July 25--30, 2020}{Virtual Event, China}
\acmBooktitle{Proceedings of the 43rd International ACM SIGIR Conference on Research and Development in Information Retrieval (SIGIR '20), July 25--30, 2020, Virtual Event, China}
\acmPrice{15.00}
\acmDOI{10.1145/3397271.3401313}
\acmISBN{978-1-4503-8016-4/20/07}

\begin{document}
\fancyhead{}

\title{A Knowledge-Enhanced Recommendation Model\\ with Attribute-Level Co-Attention}\thanks{* Corresponding author.}
\author{Deqing Yang$^*$, Zengchun Song, Lvxin Xue}
%\authornote{Corresponding author.}
\email{
yangdeqing, zcsong19, lxxue19@fudan.edu.cn
}
\affiliation{
  \institution{School of Data Science, Fudan University}
  \streetaddress{No. 220 Handan Rd.}
  \city{Shanghai 200433}
  \country{China}
}

\author{Yanghua Xiao}
\email{shawyh@fudan.edu.cn}
\affiliation{
  \institution{School of Computer Science, Fudan University}
  \streetaddress{No. 220 Handan Rd.}
  \city{Shanghai 200433}
  \country{China}
}

\begin{abstract}
Deep neural networks (DNNs) have been widely employed in recommender systems including incorporating attention mechanism for performance improvement. However, most of existing attention-based models only apply item-level attention on user side, restricting the further enhancement of recommendation performance. In this paper, we propose a knowledge-enhanced recommendation model ACAM, which incorporates item attributes distilled from knowledge graphs (KGs) as side information, and is built with a co-attention mechanism on attribute-level to achieve performance gains. Specifically, each user and item in ACAM are represented by a set of attribute embeddings at first. Then, user representations and item representations are augmented simultaneously through capturing the correlations between different attributes by a co-attention module. %Furthermore, we incorporate the objective of knowledge graph embedding into ACAM’s synthetic loss to learn model parameters better. 
Our extensive experiments over two realistic datasets show that the user representations and item representations augmented by attribute-level co-attention gain ACAM's superiority over the state-of-the-art deep models.
\end{abstract}
\vspace{-0.2cm}
\vspace{-0.2cm}
\maketitle
\vspace{-0.2cm}
{\fontsize{8pt}{8pt} \selectfont
\textbf{ACM Reference Format:}\\
Deqing Yang, Zengchun Song, Lvxin Xue and Yanghua Xiao. 2020. A Knowledge-Enhanced Recommendation Model with Attribute-Level Co-Attention. In \emph{Proceedings of the 43rd International ACM SIGIR Conference on Research and Development in Information Retrieval (SIGIR’20), July 25–30, 2020, Virtual Event, China.} ACM, New York, NY, USA, 4 pages.\\ https://doi.org/10.1145/3397271.3401313
}
%\vspace{-0.1cm}
\section{Introduction}
Encouraged by the success of deep neural networks (DNNs) in computer vision, image and natural language processing (NLP) ect., many researchers also imported DNNs to improve recommender systems. In these deep recommendation models, \emph{attention} mechanism has also been employed broadly for recommendation performance gains \cite{NAIS,DIN,SARec,DKR}. Although these attention-based models have been proven effective, the following problems restrict the further enhancement of recommendation performance. First, some of them \cite{NAIS,SARec} only employ coarse attention on item-level, i.e., each item is directly represented by a single embedding based on which user presentations are generated. Such coarse-grained embeddings can not represent users and items thoroughly. Second, although some models \cite{DIN,DKR} incorporate item features (attributes), also known as item \emph{knowledge}, to improve the expressive ability of user/item representations, they only apply attention mechanism on user side.
  
Given these problems, we propose a novel deep recommendation model with attribute-level co-attention, namely \emph{ACAM} (Attribute-level Co-Attention Model). ACAM demonstrates superior performance due to the following merits. First, its item representations and user representations are generated based on a set of \emph{attribute embeddings} rather than a single embedding, where the attributes are distilled from open knowledge graphs (KGs) as side information. Second, the co-attention module in ACAM captures the correlations between different attribute embeddings to augment user/item representations. As we know, there may exist correlations between different item attributes, indicating the latent relationships between items. For example, in movie attributes, actor \emph{Stallone} is more correlated to genre \emph{action film}, and actor \emph{GONG Li} is more correlated to director \emph{ZHANG Yimou}. Therefore, the latent relationships between the target users and the candidate items are uncovered precisely by the user/item representations augmented based on attribute correlations, resulting in enhanced recommendation performance. Furthermore, we add an objective of knowledge graph embedding (KGE) into the loss function to lean better attribute embeddings.

%In fact, there exist the correlations of different extent between different item attributes. For example, in movie attributes, actor \emph{Stallone} is more correlated to genre \emph{action film}, and actor \emph{GONG Li} is more correlated to director \emph{ZHANG Yimou}. Such attribute-level correlations also navigate us to fully exploit the latent relationships between different items, and between users and items. %In ACAM, a symmetrical neural architecture is built to refine the representations of the target user and the candidate item simultaneously. Specifically, the attribute embeddings of the item are used to guide the refinement of the user's representation, and vise versa. In other words, ACAM employs attribute-level co-attention on both user side and item side.

In summary, we have the following contributions in this paper:

1. We propose an attribute-level co-attention mechanism in a deep recommendation model, to capture the correlations between different user/item attributes sophisticatedly, and then augment user/item representations simultaneously which are helpful for recommendation performance enhancement.

%2. We verify that incorporating the objective of KGE into the learning of recommendation model results in better attribute embeddings, and thus enhanced recommendation performance is gained.

2. Our extensive experiments demonstrate our model's superiority over some state-of-the-art deep models including previous attention-based recommendation models, which apparently justify the effectiveness of incorporating attribute embeddings and employing attribute-level co-attention to co-augment user representations and item representations.

\vspace{-0.2cm}
\section{Model Description}
The task addressed by ACAM is top-n recommendation of implicit feedback \cite{NCF,NAIS}. To generate the top-n recommendation list, the target $u$ should be coupled with each candidate item $v$ and input into the model to compute $\hat{y}_{uv}$, which quantifies the probability that $u$ likes $v$, i.e., $u$ has a positive feedback to $v$. 

\begin{figure}[!htb]
%\vspace{-0.2cm}
	\center
	%\hspace{-0.2cm}
	\includegraphics[width=3in]{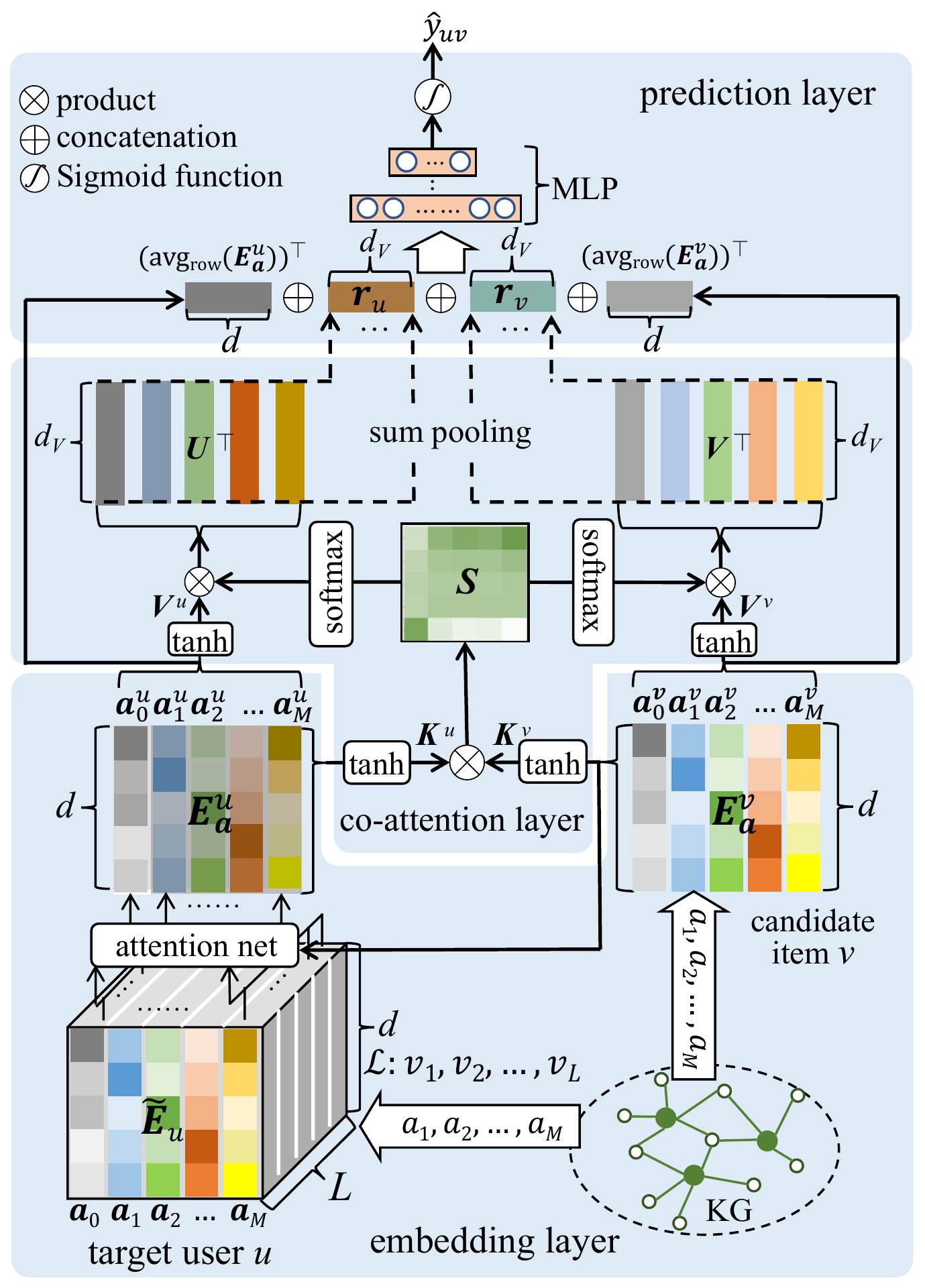}
	\vspace{-0.4cm}
	\caption{The proposed model's framework.}\label{fig:frame}
	\vspace{-0.3cm}
\end{figure}

As shown in Fig. \ref{fig:frame}, ACAM can be divided into three layers, i.e., embedding layer, co-attention layer and prediction layer. In the embedding layer, both $u$ and $v$ are represented by a representation matrix on attribute-level, rather than a single vector (embedding) as previous models \cite{NCF,NAIS,SARec}. Then, $u$'s representation and $v$'s representation are co-augmented based on the correlations (attentions) between different attributes captured by an attribute-level co-attention module. In the last prediction layer, a multi-layer perceptron (MLP) is built and fed with $u$'s representation and $v$'s representation to compute $\hat{y}_{uv}$. %Without loss of generality, we introduce the details of our model w.r.t. the task of movie recommendation in the following texts.

\vspace{-0.1cm}
\subsection{Embedding Layer} 
%\vspace{-0.1cm}
\subsubsection{Generating Item Representation}
%As we know, each item has a group of attributes (knowledge) representing its various characteristics. For example, a movie has the attributes of actor, director, genre and etc., and a song has the attributes of singer, album, composer and etc. These attributes indicate the latent relationships between different items. Therefore, in ACAM's embedding layer, we use a set of attribute embeddings to represent an item. In order to let the model based on these embeddings converge faster, all embeddings are pre-trained at first. To this end, we first construct an HIN consisting of items and their attribute entities. Then, we apply Metapath2Vec\footnote{We also tried other pre-training models including GNN-based and Tanslation-based models which are not better than Metapath2Vec apparently but time-consuming.} on this network to learn an embedding for each item and attribute entity. According to Metapath2Vec's principle, two items with common attributes have approximate embeddings. Similarly, two attribute entities (e.g., two actors) also have approximate embeddings if they are shared by some items.

In general, most items and their attributes (e.g., the actor and director of movies) in an open domain can be fetched from large-scale KGs, which constitute knowledge triplets formed as $<h,r,t>$. The head entity ($h$) is just an item, the relation ($r$) corresponds to an attribute and the tail entity ($t$) is an attribute value. For example, a triplet $<Rocky, starred, Stallone>$ describes that \emph{Stallone} is an actor of movie \emph{Rocky}. The shared attribute values well indicate the latent relationships between different items \cite{DIN,DKR}. 
Therefore, $M$ significant attributes (values) $\{a_1,a_2,...,a_M\}$ are first selected to represent an item in ACAM. The $a_i$'s embedding is denoted as $\boldsymbol{a}_i\in\mathbb{R}^{d}$ ($1\leq i\leq M$). We further use an item embedding (head entity embedding), denoted as $\boldsymbol{a}_0$, to supplement the representation of an item. For the convenience of the following introduction, $\boldsymbol{a}_0$ is also regarded as an attribute embedding. Thus, an item's representation is enriched into a matrix of $(M+1)\times d$, as shown in Fig. \ref{fig:frame}. 
Please note that an attribute may have multiple values corresponding to different tail entities in KGs. For example, a movie generally involves many actors and each actor corresponds to an entity in KGs and has a unique embedding. Thus, for an attribute with multiple values, we average all of its value embeddings (tail entity embeddings) as its attribute embedding. %Item embeddings and attribute value embeddings are both learned by a KGE objective based on transE \cite{transE}, which will be introduced in ACAM's loss function. %For those attributes having too many values, we only focus the significant ones. For example, in our experiments of movie recommendation, we only consider the top 4 actors in a movie's cast list.

\vspace{-0.1cm}
\subsubsection{Generating User Representation}
A user in ACAM is represented by the recent $L$ items that he/she has interacted with, denoted as $\mathcal{L}=\{v_1, v_2, …, v_L\}$. As the candidate item's representation, each interacted item $v_j (1\leq j\leq L)$ in $\mathcal{L}$ is also represented by the union of its $M+1$ attribute embeddings. Therefore, the target user $u$'s representation is enriched to be a cube (tensor) of $(M+1)\times L\times d$, denoted as $\boldsymbol{\tilde{E}}_u$. For those users with less than $L$ historical interactions, we fill paddings in their representations.

Next, we need to extract the features of $\boldsymbol{\tilde{E}}_u$ and compress it into a matrix. It is not only to reduce trainable parameters but also to feed the co-attention layer (module) conveniently. To this end, we aggregate the $i$-th attribute embeddings of $u$'s historical items as $u$'s $i$-th attribute embedding through an attention network. Specifically, we adopt a weighted sum pooling as
\vspace{-0.1cm}
\begin{equation}\label{eq:att}
\vspace{-0.1cm}
\boldsymbol{a}^u_i=\sum\limits_{j=1}^L  FFN(\boldsymbol{a}^j_i\oplus\boldsymbol{a}^v_i)\boldsymbol{a}^j_i
\end{equation}
where $\boldsymbol{a}^u_i$ is regarded as $u$'s $i$-th ($0\leq i\leq M$) attribute embedding, and $FFN(\boldsymbol{a}^j_i\oplus\boldsymbol{a}^v_i)$ is the computation of a feed-forward network fed with the concatenation of item's attribute embedding $\boldsymbol{a}^j_i$ and $\boldsymbol{a}^v_i$. Thus, $u$'s representation consists of $M+1$ attribute embeddings computed by Eq. \ref{eq:att}. Such user representations vary given different candidate recommended items, which have been proven more helpful for recommendation performance gains than fixed user representations \cite{DKR,NAIS}. %At last, all $\boldsymbol{a}_i^u$s and $\boldsymbol{a}_i^v$s are used together as the input of the next co-attention layer. 

\vspace{-0.1cm}
\subsection{Co-attention Layer}
In this layer, both $u$'s and $v$'s attribute-based representations are simultaneously augmented by a co-attention module of symmetrical neural architecture, as shown in Fig. \ref{fig:frame}. 

Specifically, each of $u$'s attribute embeddings, i.e., $\boldsymbol{a}_i^u$, is adjusted as the weighted sum of all $u$'s attribute embeddings (denoted as $\boldsymbol{a}_j^u$) where the weight (attention) of $\boldsymbol{a}_j^u$ is computed based on the correlation between $\boldsymbol{a}_j^u$ and $\boldsymbol{a}_i^v$. To adjust $v$'s attribute embedding $\boldsymbol{a}_i^v$, we just use the symmetric operation. Such adjustment makes the embeddings of two correlated attributes more closer to each other. As a result, the user/item representations generated based on such adjusted attribute embeddings induce more precise recommendations. For example, it makes a movie starring \emph{Stallone} easier to be recommended to a user who have watched many action films.

Formally, we first denote $u$'s representations and $v$'s representations respectively as
\vspace{-0.1cm}
$$
\vspace{-0.1cm}
\boldsymbol{E}_a^u=[\boldsymbol{a}_0^u,\boldsymbol{a}_1^u,...,\boldsymbol{a}_M^u],\quad
\boldsymbol{E}_a^v=[\boldsymbol{a}_0^v,\boldsymbol{a}_1^v,...,\boldsymbol{a}_M^v]
$$
Then, we take nonlinear transformation to obtain the key matrices $\boldsymbol{K}^u, \boldsymbol{K}^v\in \mathbb{R}^{(M+1)\times d_K}$ and value matrices $\boldsymbol{V}^u,\boldsymbol{V}^v\in\mathbb{R}^{(M+1)\times d_V}$ based on $\boldsymbol{E}_a^u$ and $\boldsymbol{E}_a^v$ as
\vspace{-0.1cm}
\begin{equation}
\vspace{-0.1cm}
\begin{split}
&\boldsymbol{K}^u=\text{tanh}(\boldsymbol{E}_a^{u\top} \boldsymbol{W}^u_K+\boldsymbol{b}^u_K),  \boldsymbol{V}^u=\text{tanh}(\boldsymbol{E}_a^{u\top}\boldsymbol{W}^u_V+\boldsymbol{b}^u_V)\\
&\boldsymbol{K}^v=\text{tanh}(\boldsymbol{E}_a^{v\top} \boldsymbol{W}^v_K+\boldsymbol{b}^v_K),  \boldsymbol{V}^v=\text{tanh}(\boldsymbol{E}_a^{v\top}\boldsymbol{W}^v_V+\boldsymbol{b}^v_V)
\end{split}
\end{equation}
where $\boldsymbol{W}^u_K, \boldsymbol{W}^v_K\in \mathbb{R}^{d \times d_K}, \boldsymbol{W}^u_V, \boldsymbol{W}^v_V\in \mathbb{R}^{d \times d_V}$ are transformation weight matrices, and $\boldsymbol{b}^u_K, \boldsymbol{b}^v_K \in \mathbb{R}^{d_K}, \boldsymbol{b}^u_V, \boldsymbol{b}^v_V \in \mathbb{R}^{d_V}$ are transformation bias vectors. Next, we obtain a co-attention map
$
\boldsymbol{S} = \boldsymbol{K}^u \boldsymbol{K}^{v\top}
$, which is a square matrix of $(M+1)\times (M+1)$ and each entry $\boldsymbol{S}_{ij}$ quantifies the affinity between $u$'s $i$-th attribute and $v$'s $j$-th attribute, i.e., the correlation between $\boldsymbol{a}_i^u$ and $\boldsymbol{a}_j^v$. Accordingly, $\boldsymbol{S}$ stores attribute-level attentions.

Based on $\boldsymbol{V}^u,\boldsymbol{V}^v$ along with $\boldsymbol{S}$, all $u$'s and $v$'s representations are revised as
\vspace{-0.1cm}
\begin{equation}\label{eq:U1}
\vspace{-0.1cm}
\boldsymbol{U} = \text{softmax}_{col}(\boldsymbol{S})^{\top} \boldsymbol{V}^u, \boldsymbol{V} = \text{softmax}_{row}(\boldsymbol{S}) \boldsymbol{V}^v
\end{equation}
where $\boldsymbol{U}, \boldsymbol{V}\in\mathbb{R}^{(M+1)\times d_V}$, and each row of them represents an adjusted attribute embedding. And $\text{softmax}_{col}(\cdot)$ and $\text{softmax}_{row}(\cdot)$ represent the softmax computation in terms of column and row, respectively. %Above formulas imply that, $v$'s attribute embeddings are used to guide the augmentation of $u$'s representation, and vise versa. 

In order to reduce the number of trainable parameters in ACAM, we set $\boldsymbol{K}^u=\boldsymbol{V}^u, \boldsymbol{K}^v=\boldsymbol{V}^v,d=d_K=d_V$ in our experiments. It has been proven that such reduction does not affect the final recommendation performance. 

The last operation in this layer is to use sum pooling\footnote{It was proven that sum pooling is a bit better than max/min/avg pooling through our experiments.} in terms of column to compress matrices $\boldsymbol{U},\boldsymbol{V}$ into the final representations of $u$ and $v$ as
\vspace{-0.1cm}
\begin{equation}\label{eq:U2}
\vspace{-0.1cm}
\boldsymbol{r}_u = \text{sum}_{col} (\boldsymbol{U}), \quad
\boldsymbol{r}_v =  \text{sum}_{col} (\boldsymbol{V})
\end{equation}

In ACAM's prediction layer, the final score $\hat{y}_{uv}$ is computed through an MLP of three layers fed with the concatenation of $\boldsymbol{r}_u$, $\boldsymbol{r}_v$, $(\text{avg}_{row}(\boldsymbol{E}_a^u))^\top$ and $(\text{avg}_{row}(\boldsymbol{E}_a^v))^\top$, where $\text{avg}_{row}(\cdot)$ is the average operation in terms of row.

\vspace{-0.1cm}
\subsection{Model Training}
As we introduced before, the item embedding $\boldsymbol{a}_0$ and attribute embedding $\boldsymbol{a}_i(1\leq i \leq M)$ are the basis of computing $\hat{y}_{uv}$. Besides the cross-entropy loss as in \cite{NCF,NAIS}, we further use a KGE objective to learn $\boldsymbol{a}_0$ and $\boldsymbol{a}_i$ better, since an item and an attribute value correspond to a head entity and a tail entity in a KG, respectively. Specifically, we adopt the objective of transH \cite{transH} model since it learns many-to-many relations effectively. Therefore, we minimize the following objective function to learn ACAM's parameters: 
\vspace{-0.1cm}
\begin{equation}\label{eq:loss}
\small
\vspace{-0.1cm}
\begin{split}
&\mathcal{O}=-\sum\limits_{(u,v)\in\mathcal{Y}}\big[y_{uv}\log\hat{y}_{uv}
+(1-y_{uv})\log(1-\hat{y}_{uv})\big]+\\
 & \lambda_1\sum_{<h,r,t >\in \mathcal{K}} \Vert (\boldsymbol{h}-\boldsymbol{w}_r^\top\boldsymbol{hw}_r)+\boldsymbol{d}_r-(\boldsymbol{t}-\boldsymbol{w}_r^\top\boldsymbol{tw}_r)\Vert _2^2  +
\lambda_2 \Vert\Theta\Vert^{2}
\end{split}
\vspace{-0.1cm}
\end{equation}
where $\mathcal{Y}$ is the union of observed user-item interactions and the negative feedbacks, and $\mathcal{K}$ is the observed triplet set in the KG. The head entity's embedding $\boldsymbol{h}$ and the tail entity's embedding $\boldsymbol{t}$ in the second term are used as $\boldsymbol{a}_0$ and $\boldsymbol{a}_i$, respectively. And $\boldsymbol{w}_r$ and $\boldsymbol{d}_r$ are the hyperplane and translation vector in transH, respectively.

\vspace{-0.1cm}
\section{Model Evaluation}
\vspace{-0.1cm}
\subsection{Experiment Settings}
%\vspace{-0.1cm}
\subsubsection{Dataset Description}
We conducted our experiments against two realistic datasets, i.e, Douban movies and NetEase songs\footnote{Douban: https://movie.douban.com, NetEase: https://music.163.com}. The statistics of these two datasets are listed in Table \ref{tab:stat}. 
%More than 900,000 ratings of 4,969 users on nearly 42,000 movies were randomly collected from Douban website. 
 We fetched Douban movies' attribute values from a large-scale Chinese KG CN-DBpedia \cite{CNDB}. In our experiments, we selected four significant attributes ($M=4$) for the two datasets, i.e., actor, director, writer and genre for Douban movies, and singer, album, composer and lyricist for NetEase songs. To reproduce our experiment results conveniently, we have published our datasets and ACAM's source code on \url{https://github.com/DeqingYang/ACAM-model}.

\begin{table}[!htb]
\vspace{-0.2cm}
    \centering
 %\small
    \caption{Statistics of the two experiment datasets.}  \label{tab:stat}
    \vspace{-0.4cm}
    \begin{tabular}{|c c c c |}
    \hline
     dataset   & user number & item number & interaction number
     \\
    \hline
       Douban  &  4,965 & 41,785 & 958,425  
       \\
       NetEase &  115,995 & 19,981 & 2,399,638 
       \\  
    \hline
    \end{tabular}
\vspace{-0.2cm}
\end{table}

%In order to achieve the recommendation of $L=10$, we first filtered out the users having 21 interaction records at least in the two datasets. 
For each user, we truncated his/her recent 10 interactions as the positive samples in test set, and the rest interactions as the positive samples in training set. We also used \emph{negative sampling} \cite{NCF} to collect negative samples for each user. Note that we inclined to those popular items (with high rating scores or more reviews) when selecting negative samples in random, to avoid such cases that a user did not rate/review an item just due to the unawareness of the item. In both model training and prediction, each positive sample was paired with 4 negative samples, which is the general setting in previous models \cite{NCF,NAIS}.

\subsubsection{Compared Models}

\textbf{NCF} \cite{NCF}: This is a DNN-based recommendation model consisting of a GMF
(generalized matrix factorization) layer and an MLP, where each user and item is represented only by a single embedding.

\textbf{NAIS} \cite{NAIS}: This is an attention-based recommendation model in which only user representations are refined by attention mechanism, and each item is represented by a single embedding.

\textbf{AFM} \cite{AFM}: It is a neural version of FM which adds an attention-based pooling layer after the pairwise feature interaction layer.

\textbf{FDSA} \cite{FSA}: This sequential recommendation model also incorporates feature-level representations but uses self-attention mechanism to only refine user representations rather than item representations. 

\textbf{RippleNet} \cite{ripple}: It is a representative KG-based recommendation model, which was compared with ACAM to highlight ACAM's strengths in knowledge (attribute) exploitation.

\textbf{DIN} \cite{DIN}: It also imports various features to enrich user/item representations. Furthermore, it uses the attention mechanism similar to NAIS to adjust user representations only.
 
%\textbf{ACAM$^-$}: This is a variant of ACAM which is fed with the attribute embeddings initialized in random instead of the ones pre-trained by Metapath2Vec. We compared ACAM with ACAM$^-$ to justify the significance of pre-training attribute embeddings by Metapath2Vec.

%\vspace{0.2cm}
In the following display of experiment results, we adopt three popular metrics evaluating top-n recommendation or ranking, i.e., HR@n (Hit Ratio), nDCG@n (Normailzed Discounted Cumulative Gain) and RR (Reciprocal Rank). To avoid statistics bias, all model's performance are reported as the average scores of 3 runnings. All baselines' hyper-parameters were set to the optimal values in their origin papers.

\begin{table*}[!htb]
\vspace{-0.2cm}
\small
\begin{center}
\caption{All models' top3/5/10 performance for the two recommendation tasks.}\label{tab:res}
\vspace{-0.4cm}
\begin{tabular}
{|p{5pt}|p{26pt}|p{20pt}|p{24pt}|p{20pt}|p{25pt}|p{22pt}|p{26pt}|p{22pt}|p{20pt}|p{24pt}|p{20pt}|p{25pt}|p{22pt}|p{26pt}|p{22pt}|}
\hline
\multirow{2}{*}{$L$}&\multirow{2}{*}{\textbf{Model}}& \multicolumn{7}{c|}{Douban movie} &\multicolumn{7}{c|}{NetEase song}\\
\cline{3-16} 
& & HR@3 & \footnotesize{nDCG@3} &HR@5 & \footnotesize{nDCG@5} & HR@10& \footnotesize{nDCG@10} &RR  & HR@3 & \footnotesize{nDCG@3} &HR@5 & \footnotesize{nDCG@5} & HR@10& \footnotesize{nDCG@10} & RR\\
\hline\hline
\multirow{8}{*}{3}&NCF & 0.8417 & 0.8483 & 0.7998  & 0.8113 & 0.7074 & 0.7492 & 0.9279 & 0.7903 & 0.7984 & 0.7689 &0.7723 & 0.6893 & 0.7193 & 0.8952\\
\cline{2-16} 
&NAIS & 0.8443 & 0.8531 & 0.8112 & 0.8253 &  0.6768 & 0.7291 & 0.9336 & 0.7998 & 0.8027 & 0.7772 & 0.7824 &  0.6747 & 0.7140 & 0.8963\\
\cline{2-16} 
&AFM & 0.8399 & 0.8455 & 0.8080 & 0.8220 &  0.7091 & 0.7499 & 0.9228 & 0.8304 & 0.8358 & 0.8112 & 0.8104 &  \textbf{0.7405} & 0.7708 & 0.9194\\
\cline{2-16} 
&FDSA & 0.8625 & 0.8690 & 0.8257 & 0.8437 &  \textbf{0.7208} & {0.7598} & 0.9394 & 0.8101 & 0.8119 & 0.7949 & 0.8011 &  0.7339 & 0.7574 & 0.8994\\
\cline{2-16} 
&\footnotesize{RippleNet} &  0.7966 & 0.8012 & 0.7694 & 0.7743 &  0.6588 & 0.7009 & 0.8957 & 0.8025 & 0.8042 & 0.7848 & 0.7901  &  0.7185 & 0.7437 & 0.8951\\
\cline{2-16} 
&DIN & 0.8322 & 0.8407 & 0.7982 & 0.8023 &  0.6453 & 0.7028 & 0.9234 & 0.7892 & 0.7936 & 0.7658 & 0.7704 &  0.6723 & 0.6950 & 0.8949\\
\cline{2-16} 
%&ACAM$^-$ & \textbf{0.8687} & \textbf{0.8765} & 0.7111  & \textbf{0.7603} & \textbf{0.9449} & \textbf{0.8348}  & \textbf{0.8399} & \textbf{0.7311} & \textbf{0.7638} & \textbf{0.9193}\\
%\cline{2-12} 
&ACAM & \textbf{0.8680} & \textbf{0.8737} & \textbf{0.8324} & \textbf{0.8477} &  0.7137 & \textbf{0.7613} & \textbf{0.9495} & \textbf{0.8541} & \textbf{0.8576}  & \textbf{0.8267} & \textbf{0.8377} &  {0.7379} & \textbf{0.7733} & \textbf{0.9301}\\	 	  	 	 	 	  	 
\hline\hline

\multirow{8}{*}{10}&NCF & 0.8335 & 0.8105 & 0.8011 & 0.8165 &  0.7003 & 0.7449 & 0.8491 & 0.7929 & 0.7875 & 0.7654 &0.7785  &  0.6980 & 0.7177 & 0.8694 \\
\cline{2-16} 
&NAIS & 0.8595  & 0.8694  & 0.8313 & 0.8440 &  0.6875  & 0.7417  & 0.9449  & 0.8026 & 0.8051 & 0.7823 & 0.7844 & 0.6774 & 0.7165 & 0.8971 \\
\cline{2-16} 
&AFM & 0.8246 & 0.8306 & 0.8041 & 0.8105 &  0.7015 & 0.7404 & 0.9163 & 0.8289 & 0.8343 & 0.8073 & 0.8090 & 0.7410 & 0.7692 & 0.9186 \\
\cline{2-16} 
&FDSA & 0.8588 & 0.8644 & 0.8292 & 0.8425 &  \textbf{0.7203} & {0.7629} & 0.9353 & 0.8325 & 0.8278 & 0.8184 & 0.8257 &  \textbf{0.7516} & {0.7725} & 0.9178 \\
\cline{2-16} 
&\footnotesize{RippleNet} & 0.8173 & 0.8224 & 0.7964  & 0.8002 &  0.6726 & 0.7172 & 0.9104 & 0.7894 & 0.7921 & 0.7670 & 0.7704 & 0.7116 & 0.7359 & 0.8904 \\
\cline{2-16} 
&DIN & 0.8325 & 0.8335 & 0.8002  & 0.8047 &  0.6579 & 0.7098 & 0.9105 & 0.7865 & 0.7906 & 0.7723 & 0.7796 &  0.6812 & 0.7065 & 0.8994 \\
\cline{2-16} 
%&ACAM$^-$ & \textbf{0.8701} &\textbf{0.8773} & \textbf{0.7137} & \textbf{0.7624} & \textbf{0.9456}  & \textbf{0.8501} & \textbf{0.8565} & 0.7402 & \textbf{0.7751} & \textbf{0.9344} \\
%\cline{2-12} 
&ACAM & \textbf{0.8682} & \textbf{0.8739} & \textbf{0.8325} & \textbf{0.8478}  &  {0.7139} &\textbf{0.7634} & \textbf{0.9504} & \textbf{0.8615} & \textbf{0.8642} & \textbf{0.8305} & \textbf{0.8423} &  0.7498 & \textbf{0.7802} & \textbf{0.9317}\\
\hline
\end{tabular}
\end{center}
\vspace{-0.2cm}
\end{table*}

\vspace{-0.1cm}
\subsection{Evaluation Results} 
%\vspace{-0.1cm}
\subsubsection{Hyper-parameter Sensitivity}
At first, we list the settings of some important hyper-parameters in our experiments in Table \ref{tab:para}. The results of hyper-parameter tuning are not displayed due to space limitation. %The influence of attribute number ($M$) depends on the significance of the incorporated attributes. 
ACAM and other models incorporating attributes enhance their performance a little when more significant attributes are fed, but ACAM still keeps its superiority. In Table \ref{tab:res}, we only display the results of $L=3$ and $L=10$ towards the two recommendation tasks. We did not focus on the scenario of $L=1$ because a user's preference can not be inferred precisely by only one historical interacted item. Given that many users' preferences may vary as time elapses, using too many historical items to represent a user would induce noises, so we neglected the cases of $L>10$. We also find that almost all models only improve their performance a bit when $L$ increases from 3 to 10. It implies that using 3 recent historical items to represent a user is adequate for many models to generate precise results. In addition, ACAM achieves the best performance when $\lambda_1$ is small, implying that too large weight of KGE objective will bias the synthetic objective of Eq. \ref{eq:loss}.

\begin{table}[!htb]
\vspace{-0.2cm}
   \centering
%\small
    \caption{Hyper-parameter settings.}  \label{tab:para}
    \vspace{-0.4cm}
    \begin{tabular}{|c c c c c c|}
    \hline
  dataset &   $d/d_K/d_V$   & $M$ & $L$ & $\lambda_1$ & $\lambda_2$  \\
    \hline
     Douban &  512  &  4 & 3,10 & 0.1 & 0.001 \\
     NetEase &  512  &  4 & 3,10 & 0.05 & 0.001 \\
    \hline
    \end{tabular}
\vspace{-0.2cm}
\end{table}

\vspace{-0.1cm}
\subsubsection{Recommendation Performance Comparison}
Table \ref{tab:res} displays all compared models' top3/5/10 recommendation performance on the two datasets. We find that ACAM outperforms NCF from the table, which justifies the effectiveness of incorporating attribute embeddings. ACAM's superiority over NAIS shows that the fine-grained user/item representations based on attribute embeddings can improve the effectiveness of attention-based models. It also justifies the rationale of augmenting item representations and user representations simultaneously by co-attention mechanism. Although AFM also captures the interactions (correlations) between different features (attributes) through attention mechanism, it does not perform well as ACAM. It shows that ACAM's co-attention mechanism captures the correlations between different attributes better than AFM's attention in terms of recommendation performance. Although FDSA and DIN also incorporate feature embeddings to enrich user/item representations, they do not perform well as ACAM, implying that co-refining user representations and item representations by co-attention mechanism is more effective than only refining user representations by attention mechanism (DIN) or self-attention mechanism (FDSA). Although RippleNet is a state-of-the-art KG-based recommendation model, it is also inferior than ACAM, showing that ACAM's co-attention mechanism exploits item knowledge (attributes) more effectively.

\vspace{-0.1cm}
\section{Related Work and Conclusion}
NCF \cite{NCF} is a pioneer work of employing DNNs into recommender systems which fuses a GMF and an MLP together to learn the user/item representation. Inspired by attention's success in computer vision, image and NLP, many researchers have also employed all kinds of attention mechanisms in recommendation models to capture diverse user preferences more precisely. For example, AFM \cite{AFM} adds an attention-based pooling layer after the pairwise feature interaction layer, to achieve content-aware recommendation. As our model, DIN \cite{DIN} also imports user/item features and introduces a local activation unit to learn user representations adaptively w.r.t. different candidate items. Similarly, NAIS \cite{NAIS} assigns different attentions to each historical item of a user to generate adaptive user representations which bring better recommendation results. FDSA \cite{FSA} applies self-attention mechanism to refine user representations which are generated based on item features.

We propose a novel recommendation model ACAM in this paper, which first represents users and items with fine-grained attribute embeddings, and then augments user representations and item representations simultaneously by an attribute-level co-attention module. Such augmented representations are proven beneficial to performance gains through our extensive experiments.

%\vspace{-0.2cm}
\bibliographystyle{ACM-Reference-Format}
\bibliography{refer}
\end{document}